\documentstyle[aps,prl,graphicx,floats]{revtex}
\newcommand{\beq}{\begin{equation}}
\newcommand{\eeq}{\end{equation}}

\begin{document}

\baselineskip=18pt

\begin{center}
{\Large\bf Numerical comparison of two approaches for the study of
           phase transitions in small systems}

\vskip 1.1cm
{\bf Nelson A. Alves\footnote{E-mail: alves@quark.ffclrp.usp.br},~~
     Jeaneti P.N. Ferrite\footnote{E-mail: jeaneti@dfm.ffclrp.usp.br}}
\vskip 0.1cm
{\it Departamento de F\'{\i}sica e Matem\'atica, FFCLRP 
     Universidade de S\~ao Paulo. Av. Bandeirantes 3900. \\
     CEP 014040-901 \, Ribeir\~ao Preto, SP, Brazil}
\vskip 0.3cm
{\bf Ulrich H.E. Hansmann \footnote{E-mail: hansmann@mtu.edu}}
\vskip 0.1cm
{\it Department of Physics, Michigan Technological University,
         Houghton, MI 49931-1291, USA}

\vskip 0.3cm
\today
\vskip 0.4cm
\end{center}
\begin{abstract}
We compare two recently proposed methods for the characterization of
phase transitions in small systems. The validity and usefulness 
of these approaches are studied for the case of the $q=4$ and $q=5$
Potts model, i.e. systems where a thermodynamic limit  and
exact results  exist. Guided by this analysis we discuss then the
helix-coil transition in polyalanine, an example of structural 
transitions in biological molecules.

 
\end{abstract}  
\vskip 0.1cm
{\it PACS-No.: 05.70.Fh, 05.50.+q, 64.60.-i, 87.15.He}



\nopagebreak


\section{INTRODUCTION} 
\noindent
 The study of phase changes in macroscopic systems has a long 
tradition in statistical physics.  Its basic assumption is 
that the dimensions of the macroscopic system are very large 
when compared with that of the constituting elements: 
phase transitions are well-defined only for infinite 
systems. However, there are  many phenomena in finite systems 
which  resemble phase transitions,  see, for instance, 
Refs. \cite{CSSG,Borderie,Lopez,BMH}. Because of their  importance 
for the understanding of the physics of  clusters of atoms \cite{Berry} 
or the folding of proteins and other biological molecules \cite{Anf}, 
to name only a few examples, these ``phase transitions'' in small
systems have  recently attracted renewed interest. The main 
question is  how the observed effects in small systems can be related
to true phase transitions in macroscopic (or infinite) systems. 
A few attempts were recently 
made in this direction, either through  studying
for finite and small systems the topology of curvatures of the 
entropy-density surface $s(e,n)$ \cite{Gross} in the microcanonical 
ensemble, or by exploring the density of complex zeros of their 
canonical partition function \cite{BMH,JK}.

The later approach is closer to the traditional view on phase transitions.
For infinite systems the physical features 
of a phase transition can be obtained from the distribution of the 
complex zeros of partition functions of finite systems.  
As the number of complex zeros $\beta_j = \beta_j(L)$ grows 
 with system size $L$ they will 
(for a system with no external field) pinch the positive real 
$\beta$-axis, and for large $L$ the corresponding value is 
the inverse of the physical critical temperature $T_c$. 
%
One example for the extension of these ideas to finite systems
is the classification scheme by Borrmann {\it et al.} \cite{BMH}
which explores the linear behavior for the limiting density
of zeros \cite{GR1,GR2}. Another  description was
suggested by Janke and Kenna (JK) who propose a  
new scaling relation to identify the order and  strength of a 
transition  from the behavior of small systems \cite{JK}.

In this paper we try to evaluate the usefulness and validity 
of these two approaches. For this purpose we apply them to
systems where a thermodynamic limit exists.  In order to be useful
the two approaches should  be able to identify the macroscopic 
phase transition from investigations of small system sizes.
Our test case is the 2D Potts model with $q=5$ states where the 
order of the phase transition is difficult to distinguish and which is 
therefore a challenging test for the new approaches. Simulations
of the Potts model with $q=4$ states add data for the case of a second
order phase transition. Our numerical investigations complement
earlier work in Ref.~\cite{JK} on the $q=10$ Potts 
model which has a pronounced first order phase transition. 

Guided by  our results for the Potts model we finally consider
the helix-coil transition in polyalanine. The purpose of this
investigation is to test whether the two approaches allow a
characterization of structural transition in biological molecules.


\section{PARTITION FUNCTION AND DENSITY OF COMPLEX ZEROS}
\noindent
 In the canonical ensemble a system is completely described 
 by its partition function
\beq
  Z(\beta) = \sum_E n(E) {\rm exp} (-\beta E) \, .
\eeq
Introducing variables $u = \exp (-k\beta)$ with conveniently
defined constant $k$ allows to write the partition function 
for discrete energy models as a polynominal:
\beq
  Z(u) = \sum_E n(E) u^{E}~.                            \label{z2}
\eeq
The number of complex
zeros $u_j$, ($u_j = {\rm exp}(-k \beta_j), j=1,2, ...$)
of this polynominal will grow with system size. 
In the case of a phase transition, we expect the zeros
(or at least the ones close to the real axis) condense for 
large enough system size $L$ on a single line 
\beq 
u_j = u_c + r_j\,{\rm exp} (i \varphi)~.                   \label{a1}
\eeq
As the system size $L$ increases, those zeros will move towards
the positive real $u$-axis  and the corresponding value is for large
$L$  the inverse of the physical critical temperature $T_c$.
Crucial information on phase transitions can be obtained from 
the way in which the first zero approaches the real $\beta$-axis.
However, such an analysis depends on the extrapolation
towards the infinite large system and does not allow  characterization 
of the situation in small systems. 

One possible extension of the above ideas to ``phase transitions'' in
small systems is the classification scheme by  Borrmann {\it et al.}
\cite{BMH}.
The idea behind this scheme is to use not only the first complex zero
$u_1$, but also higher zeros $u_2,u_3$ and $u_4$. 
Writing the
complex zeros  as $u_k = {\rm Re}(u_k) + i \, \tau_k$, 
where $\tau_k$ stands for ${\rm Im}(u_k)$,
the assumed
distribution of zeros on a straight line allows to define 
two parameters $\alpha_u$ and $\gamma_u$:
\beq
 \alpha_u = \frac{ {\rm ln}\,\phi(\tau_3) - {\rm ln}\,\phi(\tau_2)}
               { {\rm ln}\,\tau_3 - {\rm ln}\,\tau_2} \, ,    
\quad {\rm where} \quad
\phi(\tau_k) = \frac{1}{2} \left(\frac{1}{|{u}_k - {u}_{k-1}|}
         + \frac{1}{|{u}_{k+1} - {u}_{k}|}\right) \, ,        \label{phi}
\eeq
with $k$ labelling the first zeros, and 
\beq
\gamma_u = ({\rm Re}(u_2)-{\rm Re}(u_1))/(\tau_2 - \tau_1) \,.  \label{gamma}
\eeq
 Note, that our notation differs from that in Ref.~\cite{BMH} in that
 we define the discrete line density $\phi$ in function of the 
$u$-zeros instead of the temperature $\beta$.   
Following the classification scheme by Grossmann 
and Rosenhauer \cite{GR1,GR2},
phase transitions can now be classified according to the values of
these two parameters:
for $\alpha_u \le 0$ and $\gamma_u=0$ one has a phase transition of
first order, it is of second order if $0 < \alpha_u < 1$ and 
arbitrary $\gamma_u$, and for $\alpha_u > 1$ 
and arbitrary $\gamma_u$ one has a higher order transition. Besides 
the above parameters, $\tau_1$  plays also an important role: only 
for  $\tau_1 \rightarrow 0$ one  obtains in the thermodynamic
limit a real temperature for a phase transition.

Another extension of partition function zero analysis to small systems
is the approach by Janke and Kenna (JK) \cite{JK} which uses that  
the average cumulative density of zeros \cite{JK}
\beq
  G_L(r_j) = \frac{2j-1}{2L^d}                           \label{a2}  
\eeq
can be written in the thermodynamic limit and
for a first order transition as
\beq
G_{\infty}(r) = g_{\infty}(0)r + a r^{w+1} + \cdots \, .  \label{a3}
\eeq
Here,  the slope at the origin is related to the latent heat,
$\Delta e \propto g_{\infty}(0)$.
  Equations~(\ref{a1}) and (\ref{a3}) imply that the 
distance $r_j$ of a zero 
from its critical point can be written for  
large enough lattice sizes  as ${\rm Im}\,u_j(L)$ since
${\rm Re}\,u_j(L) \sim u_c$. Hence, in this limit 
Eq.~(\ref{a2}) and (\ref{a3}) lead   to the following 
scaling relation for the cumulative density of zeros
as an equation in $j$ and $L$,
\beq
\frac{2j -1 }{2L ^d} = a_1 ({\rm Im}\, u_j(L))^{a_2} + a_3 \,.\label{a5}
\eeq
A necessary condition for the existence of a phase transition 
is that $a_3$ is compatible with zero, else it would indicate
that the system is in a well-defined phase. The values of
the constants $a_1$ and $a_2$ then  characterize the phase transition. 
For instance, for first order transitions
should the constant $a_2$  take values $a_2 \sim 1$ for small $r$,
 and in this case the slope of this equation is
related to the latent heat through the relation \cite{JK}
\beq
  \Delta e = k\, u_c\, 2 \pi a_1  \,,                 \label{a7}
\eeq
with $u_c = {\rm exp}(-k \beta_c)$. 
On the other hand, a value of $a_2$ larger than $1$
indicates a second order transition whose specific heat
exponent is given by $\alpha = 2 -a_2$. 

The above approach was originally developed and tested for
systems with well defined first order phase transitions such as the  
2D 10-state Potts and 3D 3-state Potts model.
The obtained results agree with previous work from numerical 
simulations and partition function zeros analysis of systems up to $L=64$ 
\cite{Thesis,AlvesNB20}  ($L=36$ for the 3D case \cite{AlvesB43}).

The classification of phase transitions by Borrmann {\it et al.} 
has been tested for
 the finite Bose-Einstein condensates in a harmonic trap
and for small magnetic clusters and nuclear 
multifragmentation \cite{BMH}.

 In the following we will show that 
despite these successful tests both approaches can lead to wrong
conclusions if  applied blindly. For this purpose we study systems
where determination of the order of the phase transition is known 
to be computational difficult. We will concentrate on the 2D Potts 
model with  $q=4$ (second order transition) and $q=5$ (weak
first order transition) states since a series of exact  
results \cite{Wu} exist for both models. For instance, the critical 
temperatures are known to be $ \beta_c = {\rm ln}(1+\sqrt q)$ 
and the latent heat is given by $\Delta e(q=5) = 0.0529187$. 
Finally, we research whether the  two approaches can be used to characterize
structural transitions in biological molecules. 

\section{RESULTS FROM POTTS MODEL SIMULATIONS}
\noindent
We start by presenting our results for the two Potts 
models with $q=4$ and $q=5$  on small lattices.
 Our study was performed on lattice sizes $L=16, 32, 64$ and $128$
for $q=4$ and  $L=8, 12, 16, 20, 24, 32, 64, 96\,$ and $128$
for $q=5$.
 For these models and system sizes we have evaluated  the complex 
partition function zeros from heat-bath simulations with large 
statistics at temperatures $\beta_0$ listed in Table 1. For each 
lattice size, our results rely on 16 bins of 500000 measurements. 
For $L < 96$ each measurement was separated by an additional Monte 
Carlo sweep which was discarded; for larger sizes measurements were 
only taken every 5th sweep.

For lattice sizes up to $L=20$,
we could calculate all complex zeros from the polynominal
form of the partition function  using MATHEMATICA. For larger 
lattice sizes  such a direct evaluation 
is no longer possible with standard numerical algorithms,  
even when double precision was used. 
Only by using the scan method 
(see \cite{AlvesIJMPC} and references therein)
we were able to obtain reliable estimates for the first $J$
complex zeros. Note, that $J$  is limited by the precision of our
data: insufficient statistics of our simulation will lead to 
presence of so-called fake zeros \cite{AlvesNPB92}.
 For instance, for the $q=4$ Potts model we could only get 
reliable estimates for  the first three zeros (Table 2) 
while for the $q=5$ Potts model we were able to obtain estimates for 
the first four partition function zeros (presented in Table 3).
The error bars were calculated from the fluctuation of the averages 
taken over each of those 16 bins.

\subsection{The Borrmann Approach}
\noindent
 From the listed zeros in  Tables  2 and 3 we obtain  our first
result:  even for our smallest lattice size 
($L=16$ for $q=4$ and $L=8$ for $q=5$) the zeros stay on a straight line  
and the real parts are  approximately constants. 
This is a necessary condition for both the Borrmann and the JK approach.
We  first  calculate for both $q=4$ and $q=5$ Potts
model the parameters $\alpha_u$ and $\gamma_u$ on which the
Borrmann classification scheme relies.  Their values are listed 
in Table 4 for all lattice sizes,  and also plotted in Fig.~1a for
the $q=4$ Potts model and in Fig.~1b for the case of $q=5$. 
For the $q=4$ Potts model we were only able to obtain reliable estimates 
for the first three zeros for each lattice size. However, in order
to calculate the parameter $\alpha_u$ we need  four zeros. 
For this reason, we have included here the less reliable fourth zero
which is not listed in Table 2. As a consequence, our results for
the $q=4$ Potts model are less reliable than these for the 
$q=5$ model where we have also acceptable estimates for the fourth zeros.
Estimates and standard deviations presented in Table 4 were obtained
by means of the bootstrap method \cite{Efron} based on our
statistics of 16 independent bins for each lattice size.
Details on this method are 
presented in Appendix A. 

For both
models, the obtained values of Borrmann parameters $\alpha_u$ and 
$\gamma_u$ show no size dependence, but rather seems to fluctuate around
some average value. For the case $q=4$, the values
of $\alpha_u$ and $\gamma_u$ are compatible with the well-known fact that
this model has a second order phase transition in the thermodynamic limit.
 However, our approach seems to fail for the case $q=5$ which has a
weak first order phase transition and where we  would expect
$\alpha_u  \le 0 $ and $\gamma_u =0$. Our results  rather indicate
a second order transition ($0 < \alpha_u < 1$ and $\gamma_u$ arbitrary).
However, the  $q=5$ Potts model  is well-known to have a very large 
correlation length (of order 2000 lattice units \cite{LandauB39}), 
and hence, its true behavior may only be caught for very large lattice 
sizes. Fig.~1b does not give any indications that we are even close
to lattice sizes where the approach by Borrmann would lead to the
correct result since the values for $\alpha_u$ and $\gamma_u$ show no 
systematic size dependence for the lattice sizes which we 
have studied ($L \le 128$).

 Our results indicate  that an uncritical application of Borrmann approach
may lead to wrong conclusions on the nature of phase transition in a
system. In the case of the $q=5$ Potts model this approach fails to
identify the nature of a phase transition from the distribution of zeros
on small lattices. This seems to limit the usefulness of this method 
to systems where the order of the transition is clear.

\subsection{The JK Approach}
\noindent
A similar method to study phase transitions in small systems is the one
proposed by Janke and Kenna \cite{JK}. In this approach 
one has to calculate  the average cumulative density of zeros $G(r)$ 
from the zeros listed in Table 2 and 3.
In order to investigate  the phase transitions in our two systems
one has to fit the cumulative density $G(r)$ to
\beq
G(r) = a_1 r^{a_2} + a_3                       \,.      \label{a6}
\eeq
The aim of this fit is to obtain an estimate for $a_3$ since
$a_3=0$ indicates the existence of a phase transition.
However, a simple evaluation of this fit may 
be misleading if lattice sizes are too small. In this case it may be
necessary to study  the FSS dependence of these quantities, 
i.e. how their estimates are related to ones obtained 
for larger systems or even in the thermodynamic limit.

In the case of the
$q=4$ Potts model we have to  rely  only on the first
three zeros for each lattice size which 
is a too small number for a meaningful three-parameter fit.
For this reason, we have combined the zeros of two neighboring lattice sizes 
and our fit therefore relies on six zeros for each pair. The so obtained
estimates for $a_3(L)$ are listed in Table 5. For the
case of the $q=5$ Potts model  we have four zeros for each lattice size,
but we have also calculated   $a_3(L)$ from fits where 
the eight zeros of two neighboring
lattice sizes were combined, which leads to  a more robust estimate
of this parameter.  The so obtained sets of  $a_3(L)$
are listed in Table 6.  We note that for both models the
values of $a_3 (L)$ are for all lattice sizes compatible with zero,
demonstrating that the two Potts models have indeed a phase transition.

The next question is whether and for what sizes the above approach is 
able to identify the order of the transition. This requires to calculate
an accurate estimate for the quantity $a_2$ in Eq.~(\ref{a6}). 
For this purpose, we set $a_3$ in Eq.~(\ref{a6}) to zero and replace 
that equation by the simpler two-parameter fit
\beq
G(r) = a_1 r^{a_2}~.                    \label{a6a}
\eeq
Extracting the parameters $a_1$ and  $a_2$ from this fit requires 
an extremely careful data analysis and error estimation.
For this reason we have 
again used the  bootstrap method for estimating 
averages and standard errors.  The obtained values of $a_1$ and $a_2$
are  presented for all lattice sizes and both $q=4$ and $q=5$ 
in Tables 5 and 6. In Table 6 we list these parameters   
as a function of the first four complex zeros.

Let us  first consider the case $q=4$. 
Fig.~2 displays the  values for $a_2(L)$ as a function of 
lattice size. In this plot  we do not observe any dependency of
$a_2$ from the system size and the possibility of a first order 
transition ($a_2=1$) is clearly excluded.  
Hence, our analysis reproduces the well-known fact that the 
$q=4$ Potts model has a second order phase transition. 
However, with our estimate of $a_2$ from our largest
lattice size, we find as critical exponent $\alpha=2-a_2 = 0.430(18)$.
 This value is far from the thermodynamic limit $2/3$
 \cite{Nau,Creswick,Sokal}. This discrepancy may be due to the  
dependence of $a_2(L)$ on the higher and less precise zeros in the
fits. This dependence
is less pronounced if we merge the zeros of all lattice sizes.
Evaluating the resulting cumulative density leads to a value of
$a_2 = 1.466(23)$ which corresponds to   $\alpha = 0.534(23)$.
This value of the critical exponent $\alpha$  is closer to, but 
still far from the theoretical value. It follows that on the simulated
lattice sizes the JK scaling relations do  allow  a qualitative 
characterization of the transition in the $q=4$ Potts model, but
not to obtain   quantitative  results such as the numerical value
of critical exponents. This limitation is a somehow surprising result since
the density and the cumulative density of zeros seem to be free \cite{JK} from
the multiplicative logarithmic corrections to the leading
power-law finite-size behavior \cite{Sokal,Caselle} which hamper
the determination of the critical exponents $\alpha$ and $\nu$ in
other approaches.

At first sight, the situation seems to be worse for the $q=5$ Potts model.
Even if we discard the fourth zero, which has larger fluctuations, 
and repeat our analysis only for the more reliable first 
three ones,  our data do not  lead to the expected result  
$a_2 \sim 1$. This can also be seen from Fig.~2 where we plot also for
this model the parameter $a_2(L)$ as a function of lattice size $L$.
However, unlike in the Borrmann analysis of the $q=5$ Potts model,
our data show a clear trend towards the expected value for a first order
phase transition ($a_2 \sim 1$). 
Our data suggest, that through by an extrapolation towards 
large (ideally infinite) lattices the true value of $a_2$ can
be determined with the JK approach even for the $q=5$ Potts model.
This is consistent with the fact that the $q=5$
Potts model has a very large correlation length (of order 2000 
lattice units), and 
hence, its true behavior can only be caught 
for very large lattice sizes.  
The log-log plot of  $a_2(L)$ in Fig.~3 indicates that 
one should use a  polynominal fit for the extrapolation towards
the thermodynamic limit:
\beq
     a_2(L)= a_2(\infty) + b L^{-c}  \,.            \label{a122}
\eeq
In order to see how acceptable the expected limit $a_2=1$ is, 
 we replace the above equation by a two-parameter fit
${\rm ln}\,(a_2(L)-1) = {\rm ln}\,b - c\,{\rm ln} L$ (i.e.
we set $a_2=1$).  In this way, we obtain 
$b=0.717(16)$ and $c=0.1765(69)$, 
with the goodness of fit \cite{recipes} $Q=0.72$ for four zeros and 
$b=0.7910(88),\, c=0.2050(40)$, with a smaller $Q=0.09$ for three zeros.
The corresponding fit for the case of four zeros is  shown in Fig.~4 and
demonstrates that  our data are indeed compatible with 
the expected value $a_2 \sim 1$ for a first order phase transition.
However, we still can obtain acceptable fits ($Q > 0.7$)
for the whole range of $0 \le a_2 \le 1.013$ 
(but note that values of $a_2 < 1$ are difficult to interprete in the
JK theory, and rather indicate numerical instabilities) which shows
the difficulty in determining the
order of the phase transition for the $q=5$ Potts model.
Only if we restrict our analysis to the first three zeros the goodness of
fit shows a maximum for $a_2 \approx 1$.

The JK approach allows us also to calculate the latent heat
for the case of a first order phase transition by means of Eq.~(\ref{a7}).
However, that equation is valid only when $a_2(L) = 1$. 
In the case of the $q=5$ Potts model we have $a_2 (L) > 1$ for all
lattice sizes. Hence, we can not use Eq.~(\ref{a7}) 
to calculate the latent heat from
our values of $a_1(L)$. Instead we have replaced Eq.~(\ref{a6a}) by
\beq
 G_L(r) = A_1(L) r~,
\eeq
i.e.  $a_2(L)$ is substituted in Eq.~(\ref{a6a}) by $a_2(\infty)$, 
and have studied the
finite size scaling of the new quantity $A_1(L) $ which has the same
limiting value than $a_1(L)$.  Values for $A_1$
are  listed in Table~6.  In order to calculate the latent heat
of the $q=5$ Potts-model we evaluate firstly $a_1 = A_1(\infty)$ from the
finite-size-scaling fit
\beq
 A_1(L) = A_1(\infty) + B L^{-C}~,                    \label{fit}
\eeq
which leads with a goodness-of-fit $Q=0.68$ to the values 
$a_1=0.0284(18), B=1.381(21)$ and $C=0.6187(80)$
when the first four zeros are used for each lattice size. Restricting
the analysis to $A_1(L)$ calculated only from the first three zeros 
at each lattice size leads to $a_1=0.03085(78)$ 
(with a goodness of fit $Q=0.75$).
 Applying Eq.~(\ref{a7}) we therefore find as latent heat 
$\Delta e = 0.0551(35)$
($\Delta e = 0.0599(15)$ in case of three zeros) which is very close to the 
theoretical value $\Delta e(q=5) = 0.0529\cdots$. This result is
surprisingly good when compared with other numerical estimates. 
For instance, a recent study using lattices of up to $L=4000$ 
led to a latent heat value of $\Delta e = 0.054$ \cite{NiOk}.

Note that Eq.~(\ref{fit}) corresponds 
to the finite size scaling behavior of the 
specific heat if we identify $C = (1-\alpha)/\nu$
since the latent heat scales with $C_v\, \Delta T_c(L)$, 
and at the critical temperature we have
 $\Delta T_c(L) \sim L^{-1/ \nu}$ and
\beq
       C_v^{\rm max}(L) = c_1 + c_2 L^{\alpha/\nu}  \,.        \label{a14}
\eeq
This allows us to  identify theoretically a second correction term in 
Eq.~(\ref{fit}):
\beq
 A_1(L) = A_1(\infty) +B_1 L^{-(1-\alpha)/\nu} +B_2 L^{-1/\nu}~. \label{fit2}
\eeq
A check shows that our value of $C=0.6187(80)$  
is indeed close to $(1-\alpha)/\nu =0.627$ where the values of the
called pseudocritical exponents $\alpha=0.63(5)$ and 
$\nu=0.59(3)$ were taken from finite estimates in Ref.~\cite{LandauB39}.

  We also see from Eq.~(\ref{fit2}) that for a strong first order phase
transition (as for instance in the earlier studied $q=10$ Potts 
model \cite{JK}) where we have  $\nu =1/d$ and $\alpha =1$,
we do not have the first correction term in Eq.~(\ref{fit2}). 
 This explains why  even from small lattice sizes very good estimates 
of the latent heat 
 ($\Delta e =0.698(2)$ 
which one has to compare with the exact value for the $q=10$ Potts model: 
  $\Delta e(q=10) = 0.6960494$ \cite{Wu})
could be obtained for the $q=10$ Potts model in Ref.~\cite{JK}. 
 In fact, these estimates can be easily improved by
including a second correction term, which goes as $L^{-2}$ for this 
model towards the new value $\Delta e = 0.696(92)$. The large error 
is due to the fact that only the largest
sizes $L=32, 38, 48\, {\rm and}\, 64$, for which the first correction
term can be discarded, were considered in our calculation.
Here, our data relies on
the values quoted in Ref.~\cite{Thesis}.  

Our results indicate the JK scaling relations are more suitable
than the Borrmann approach for studying phase transitions from the
behavior of small systems. Unlike the Borrmann approach the order of
the phase transition could be determined for both the $q=4$ and the
$q=5$ Potts model. In the latter case (and for the $q=10$ Potts model
which has also a first order phase transition) it was possible to calculate
the latent heat with good accuracy. However, the approach failed
to give the correct value for the critical exponent $\alpha$ in
the case of the $q=4$ Potts model which has a second order transition.

\section{Helix-coil transition in polyalanine}
\indent
A common, ordered structure in proteins is the \hbox{$\alpha$-helix}  
and it is conjectured that formation of $\alpha$-helices is a 
key factor in the early stages of protein-folding \cite{BSG}. 
It is long known that $\alpha$-helices undergo a sharp transition 
towards a random coil state when the temperature is
increased. The characteristics of this so-called 
helix-coil transition have been studied extensively \cite{Poland}. 
In previous work \cite{HO98c,AH99b,KHC99c} evidence was presented that
 polyalanine exhibits a phase transition between the ordered
helical state and the disordered random-coil state when interactions
between all atoms in the molecule are taken into account.
Here, we revisit polyalanine and investigate the helix-coil transition
by means of the partition function analysis with the 
classification scheme of Borrmann and Janke/Kenna.

Our investigation of the helix-coil transition for polyalanine is
based on a detailed, all-atom representation of that homopolymer.
Since one can avoid the complications of electrostatic and
hydrogen-bond interactions of side chains with the solvent for alanine
(a nonpolar amino acid), explicit solvent molecules were neglected.
The interaction between the atoms  was
described by a standard force field, ECEPP/2,\cite{EC}  (as implemented
in the KONF90 program \cite{Konf}). Chains of up to $N=30$ monomers were
considered, and our results rely on multicanonical simulations \cite{MU}
of $N_{sw}$ Monte Carlo sweeps starting from a random
initial conformation, i.e. without introducing any bias.
We chose $N_{sw}$=400,000, 500,000, 1,000,000, and 3,000,000 sweeps
for $N=10$, 15, 20, and 30, respectively. 
Measurements were taken every fourth Monte Carlo sweep.
Additional  40,000 sweeps ($N=10$)
to 500,000 sweeps ($N=30$) were needed for the  weight factor 
calculations by the iterative  procedure described first in
Refs.~\cite{MU}.  In contrast to our first calculation of complex zeros
presented in Ref.~\cite{AH99b}, where we divided the energy range into
intervals of lengths 0.5 kcal/mol in order to make Eq.~(\ref{z2})
a polynomial in the variable $u=e^{-\beta/2}$, we avoided  any 
approximation scheme in the present work. 
  This because the above 
approximation works very well for the first zero, but not
for the next ones. Since we  need high precision estimates
also for the next zeros we applied again the scan method. 

In Table 7  we present our first four partition function zeros
for 7 bins, although the fourth one
is less reliable due to the presence of fake zeros.
 It is hardly possible to divide our production
data in larger number of bins due to the limited statistics of our 
runs. Using again  the bootstrap method we first calculated out of the
zeros for each bin the parameters $\alpha_u$ and $\gamma_u$ which characterize
in the Borrmann approach phase transitions in small systems. As one
can see in Table 8 the so obtained values for polyalanine are characterized
by large error bars and show no clear trend with chain length. 
It seems that the median of the $\alpha_u$ values is $\alpha_u = 0$
which would indicate a first order transition. However, our data 
have too large errors to draw such  a conclusion
 on the nature of the helix-coil transition.

For this reason, we tried instead  the JK scaling relations.
Table 9 lists the parameter $a_3(N)$ of Eq.~(\ref{a5}).  
Here, the average cumulative density of zeros is replaced by
\beq
  G_N(r_j) = \frac{2j-1}{2N}  \, ,                      \label{poly1}  
\eeq
where we have translated the linear length $L$ 
as $N^{1/d}$ ~\cite{AH99b}. Therefore all finite size
scaling relations can be written is terms of the number of monomers $N$.

The values $a_3$ are compatible with zero for chains of all length
indicating that we have indeed a
phase transition. In order to evaluate the kind of transition we also
calculate the parameters $a_1(N)$ and $a_2(N)$ which we also 
summarize in Table 9. 
Similar  to the case of the $q=5$ Potts-model, the parameter $a_2(N)$
decreases with system size. The log-log plot of this quantity as a
function of chain length in Fig. 5 suggests again a scaling relation
\beq
a_2(N) = a_2 + b N^{-c} \, .                              \label{poly2}  
\eeq
A numerical fit of our data to this function leads to 
a value of $a_2 =1.31(4)$ with $Q=0.95$. 
Using $\alpha = 2 - a_2$ we find $\alpha =0.70(4)$
which is barely compatible with our previous value of $\alpha =0.86(10)$ in 
Ref.~\cite{AH99b}, obtained from the maximum of specific heat. 
A fit of all four chain lengths can also not exclude a value $a_2 =1$
since we can find acceptable fits with $Q > 0.55$ in the 
range $ 0.92 < a_2 < 1.44 $. However,
a close examination of Fig.~5 shows that the $N=30$ data point shows a 
considerable deviation from the trend suggested by the smaller chain length.
Since the $N=30$ data are the least reliable, we also evaluated 
Eq.~(\ref{poly2}) omitting the $N=30$ chain. This leads to a value of
$a_2=1.16(1)$ and a critical exponent $\alpha=0.84(1)$ which is now 
compatible with our previous result $\alpha = 0.86(10)$.

Let us summarize our results for the helix-coil transition studies
of polyalanine. The JK approach 
is able to reproduce for polyalanine results obtained in previous 
work \cite{AH99b}, but does not lead here to an improvement over other
finite size scaling techniques. Especially, the JK approach does not allow
to establish the order of the helix-coil transition from simulation
of small chains. Our results for the parameter $a_2$ seem to favor a
second order transition, but are due to large errors and disputed by
Ref.~\cite{KHC99c} were indications for a finite latent heat were found.
In the present study we considered  only
a special kind of biological molecules, homopolymers of amino acids,
where in principle the thermodynamic limit can be considered. This allows
finite size scaling which was an essential tool for obtaining the
correct results with the JK approach. However, in general such finite
size scaling is not possible for biomolecules that have a distinct size
and composition. In these cases we have to rely on Borrmann's approach.
However, in our example of polyalanine, Borrmann's approach led to 
even less conclusive results since the error bars were large. 
Hence, it seems that both approaches are  limited in their usefulness
for the study of phase-transitions in biomolecules.

\section{Conclusion}
\indent
We have evaluated two recently proposed schemes for characterizing
phase transitions in small systems. Simulating the $q=4$ and $q=5$
Potts model, where for the thermodynamic limit we can compare our
data with exact results, we found that both Borrmann and JK approach
work well when the order of the phase transition is not in question
(such as in the case of the $q=4$ Potts model (second order) or the
earlier studied first order transition in the $q=10$ Potts model).
The situation is different for cases such as the $q=5$ Potts model where
it is difficult to distinguish between a weak first order or a strong
second order transition. A careful application of the JK approach
led to the correct result of a weak first order transition for the
$q=5$ Potts model while the Borrmann approach did not allow a correct
identification of this transition with the lattice sizes studied by us.
 Our results from Potts model simulations indicate  that both approaches
have to be applied with great care if one wants to avoid  wrong 
conclusions on the nature of phase transition in a system. 
This may limit their usefulness to systems where the order of the 
transition is clear and to systems which are not too small. Especially, 
a first application of both approaches to helix-coil transitions 
in polyalanine  demonstrates the difficulties appearing when they
are applied to the study of phase transitions in  biomolecules.

\section{Appendix: Bootstrap method} 

  Bootstrap is a simulation method based on a given
data sample, to produce statistical inferences like standard
error and bias-corrected estimator for the mean sample \cite{Efron}.
  The bootstrap algorithm consider that our data
can be obtained from an unknown probability distribution 
$F$ by random sampling
\beq
  F \rightarrow {\rm\bf X} = (x_1, x_2, \cdots , x_n) \, .   \label{p1}
\eeq
  Here the points $x_i$ refer to our $n=16$ bins 
for each lattice size in the Potts model or to $n=6$ bins
for polyalanine. For each bin, the points 
$x_i$ contain the first four complex zeros. 

  Our statistics of interest are estimates for mean values 
$\hat{\theta}=\theta({\rm\bf X})$ for the parameters 
$\alpha$ and $\gamma$ in the Borrmann approach
\begin{eqnarray}
  \hat{\alpha} = {\cal F}(\rm\bf X) \,  \\                  \label{p2}
  \hat{\gamma} = {\cal G}(\rm\bf X)                         \label{p3}
\end{eqnarray}
and their respectives standard errors ${\rm se}(\hat{\alpha})$
and ${\rm se}(\hat{\gamma})$.
  Here $\cal F$ and $\cal G$ stand for the application of 
functions in 
Eqs.~(\ref{phi})~and~(\ref{gamma}).  

The bootstrap algorithm 
follows by considering our sample as an empirical distribution
$\hat{F}$, where each data has the probability $1/n$.
  The way bootstrap assigns an accuracy to our parameters
does not depend on any theoretical calculation but
random samples of size $n$ drawn with replacement from $\hat{F}$, 
called a bootstrap sample,
\beq
 \hat{F} \rightarrow {\rm\bf X^*} = (x_1^*, x_2^*, \cdots, x_n^*) \,,
                                                          \label{p4}
\eeq
where each value $x_i^*$ equals any one of the $n$ values $x_i$.

 Therefore for each bootstrap sample ${\rm\bf X^*}$
there is a bootstrap replication of $\hat{\theta}$
which we denote as  $\hat{\theta}^*$.
 If we repeat this process, the sample standard deviation 
can be obtained from $B$ replications
\beq
   \hat{se}_B = \{ \sum_{b=1}^{B} 
       [ \hat{\theta}^*(b) - \hat{\theta}^*(\cdot)]^2/(B-1) \}^{1/2}\, ,
                                                               \label{p5} 
\eeq
where the bootstrap mean 
$\hat{\theta}^*(\cdot) = \sum_{b=1}^B \hat{\theta}^*(b)/B $.
The limit of $\hat{se}_B$ as $B$ goes to infinity gives the ideal
bootstrap estimate of $se(\hat{\theta})$. \\

 The histogram in Fig. 6 represents our $\hat{\theta}^*$ distribution
 for the parameter $\alpha$ in our simulation of the  $q=4$ Potts model
 for $L=128$ with  $B=400$ replications.
 We obtain the bootstrap mean $ \hat{\alpha}^* = 0.46(17)$.
 On the other hand, if calculating simple averages from our 
 16 measurements leads to $\hat{\alpha} = 0.41(18)$.

 The bootstrap estimate of bias based on $B$ replications
is given by 
\beq
    \hat{bias}_B =   \hat{\theta}^*(\cdot) -   \hat{\theta} \,.  \label{p6}  
\eeq
  Therefore, the bias-corrected estimator is 
$ \bar{\theta} = \hat{\theta} - \hat{bias} $.
It is also convenient to evaluate the 
ratio of estimated bias to standard deviation 
$\hat{bias}_{400}/ \hat{se}_{400}$. A small number indicates
an unneeded bias correction in face of the standard error.
 The called root mean square error of estimator
$\hat{\theta}$ for $\theta$ can be defined to take into account
both bias and standard deviation \cite{Efron},
\beq
\sqrt{E_F[(\hat{\theta}-\theta)^2]} = 
  \sqrt{se_F(\hat{\theta})^2 + bias_F(\hat{\theta},\theta)^2} \\
  \simeq se_F(\hat{\theta})\, [1+\frac{1}{2}(\frac{bias_F}{se_F})] \,.
                                                               \label{p7}  
\eeq

If we take $\hat{bias} = \hat{bias}_{400}$, our final result
is $\bar{\alpha}=0.36(18)$  for $q=4$ and $L=128$.
In the tables we quote our final estimates 
$ \bar{\theta} = \hat{\theta} - \hat{bias}$ and the above 
bias-corrected standard errors.

\ \\
\noindent
{\bf  Acknowledgements}\\
 U.H.E. Hansmann gratefully acknowledges support by research grant 
of the National Science Foundation (CHE-9981874), and 
N.A. Alves support by CNPq (Brazil). Part of this article was 
written when U.H. was visiting  the University of Central Florida.
He thanks Brian Tonner, Alfons Schulte and the other faculty at
the Department of Physics for the kind hospitality
extended to him.



\newpage

{\huge Figure Captions:} \\
\begin{description}
\item[Figure 1.]  $\alpha_u(L)$ and $\gamma_u(L)$ estimates for 
                  $q=4$ Potts model 
           data (a) and for $q=5$ Potts model data (b).\\
\item[Figure 2.] The parameter $a_2(L)$ as function of system size $L$ for 
          the $q=4$ Potts model and the $q=5$ Potts model.\\
\item[Figure 3.] The parameter $a_2(L)$ as a function of system size $L$ for
          the $q=5$ Potts model in a log-log plot.
\item[Figure 4.] $a_2(L) - 1 (=a_2(\infty))$ as a function of system size $L$
          in a log-log plot.
          The straight line through the data points is from our fit.
\item[Figure 5.] The parameter $a_2(N)$ for polyalanine molecules of 
                 length $N$ in a log-log plot.
\item[Figure 6.] Histogram of 400 bootstrap replications for
         the parameter $\alpha_u$ from four-state Potts model data.
\end{description}
\ \\
{\huge Table Captions:}\\
\begin{description}
\item[Table 1.] Heat-bath simulations at $\beta_0$ for four and 
                five-state Potts model in two dimensions.
\item[Table 2.]  Complex partition function zeros
         $u_j,\, (j=1,2 ~{\rm and}~ 3)$ for four-state Potts model.
\item[Table 3.]  Complex partition function zeros
 $u_j,\, (j=1,2,3 ~{\rm and}~ 4)$ for five-state Potts model.
\item[Table 4.]  Bootstrap bias-corrected estimates and
          bias-corrected standard error for the parameters
          $\alpha_u$ and $\gamma_u$ for $q=4$ and $q=5$.
\item[Table 5.] Bootstrap bias-corrected estimates and
          bias-corrected standard error for the JK parameters 
          from first three zeros for $q=4$.
\item[Table 6.] Bootstrap bias-corrected estimates and
          bias-corrected standard error for the JK parameters 
          from first four zeros for $q=5$.
\item[Table 7.] Partition function zeros for polyalanine.
\item[Table 8.]  Bootstrap bias-corrected estimates and
          bias-corrected standard error for the parameters
          $\alpha_u$ and $\gamma_u$ for polyalanine model.
\item[Table 9.]  Bootstrap bias-corrected estimates and
         bias-corrected standard error for the JK parameters 
         for polyalanine.
\end{description}

\newpage
\cleardoublepage


\begin{table}[ht]
\renewcommand{\tablename}{Table}
\caption{\baselineskip=0.8cm Heat-bath simulations at $\beta_0$
 for four and five-state Potts model in two dimensions.}
\begin{center}
\begin{tabular}{lll}
\\[-0.3cm]
$~L$ &$\beta_0(q=4)$ &$\beta_0(q=5)$     \\
\\[-0.35cm]
\hline
\\[-0.3cm]
~~8  &           &~1.1283\\
~12  &           &~1.1489\\
~16  &~1.084     &~1.1580\\
~20  &           &~1.1626\\
~24  &           &~1.1655\\
~32  &~1.090     &~1.16866\\
~64  &~1.096     &~1.17240\\
~96  &           &~1.17332\\
128  &~1.09755   &~1.17373\\
\end{tabular}
\end{center}
\end{table}


\begin{table}[ht]
\renewcommand{\tablename}{Table}
\caption{\baselineskip=0.8cm Complex partition function zeros 
         $u_j,\, (j=1,2 ~{\rm and}~ 3)$ for four-state Potts model.}
\begin{center}
\begin{tabular}{lclclcl}\\
\\[-0.3cm]
$~L$ &  Re$(u_1)$ &~~Im$(u_1)$      &  Re$(u_2)$  &~~Im$(u_2)$
     &  Re$(u_3)$ &~~Im$(u_3)$ \\
\\[-0.35cm]
\hline
\\[-0.3cm]
~16 &0.339010(28) & 0.016483(33)    &0.335906(66) &0.032757(84) 
    &~0.33423(40) & 0.04589(23) \\
~32 &0.335633(37) & 0.006419(20)    &0.334563(48) &0.012875(52) 
    &~0.33430(18) & 0.01787(22) \\
~64 &0.334196(10) & 0.002482(19)    &0.334034(47) &0.005050(31) 
    &~0.33359(17) & 0.006824(77) \\
128 &0.333674(10) & 0.0009474(78)   &0.333542(12) &0.001952(10) 
    &~0.33333(07) & 0.002618(28) \\
\end{tabular}
\end{center}
\end{table}


\begin{table}[ht]
\renewcommand{\tablename}{Table}
\caption{\baselineskip=0.8cm Complex partition function zeros 
 $u_j,\, (j=1,2,3 ~{\rm and}~ 4)$ for five-state Potts model.}
\begin{center}
\begin{tabular}{lclclclcl}\\
\\[-0.3cm]
$~L$ &  Re$(u_1)$   &~~Im$(u_1)$    &  Re$(u_2)$   &~~Im$(u_2)$ 
     &  Re$(u_3)$   &~~Im$(u_3)$    &  Re$(u_4)$   &~~Im$(u_4)$  \\
\\[-0.35cm]
\hline
\\[-0.3cm]
~~8  &~0.321522(26)  &0.033184(19)   &~~0.313723(58)  &0.067481(33) 
     & 0.30728(22)   &0.09619(18)    &0.3013(11)      &0.1223(11) \\
~12  &~0.316268(26)  &0.018168(14)   &~~0.312678(29)  &0.037897(29) 
     & 0.31030(15)   &0.05411(13)    &~0.30850(87)    &0.06935(54) \\
~16  &~0.313834(25)  &0.011808(10)   &~~0.311797(34)  &0.024957(25) 
     & 0.31040(13)   &0.03614(11)    &~0.30859(82)    &0.04550(54) \\
~20  &~0.312511(15)  &0.008441(14)   &~~0.311073(14)  &0.018037(18) 
     &~0.310277(77)  &0.026062(97)   &~0.30994(57)    &0.03288(29)\\
~24  &~0.311688(20)  &0.006394(11)   &~~0.310639(20)  &0.013791(20)  
     &~0.310037(80)  &0.020166(78)   &~0.31008(32)    &0.02560(27)\\
~32  &~0.310760(18)  &0.0041194(87)  &~~0.310158(13)  &0.008993(17) 
     &~0.309810(42)  &0.013193(55)   &~0.30957(18)    &0.01651(15) \\
~64  &~~~0.3096252(58)&0.0014120(24)  &~~0.3094365(61)  &0.0031623(42) 
     &~0.309353(14)  &0.004685(17)   &~~0.309370(58)  &0.005988(64) \\
~96  &~~~0.3093404(43)&0.0007373(31) &~~~0.3092477(54)&0.0016948(52)   
     &~~~0.3092213(93)&0.0025327(81) &~~0.309199(33)  &0.003198(25)\\
128  &~~~0.3092206(71)&0.0004654(28) &~~~0.3091614(67)&0.0010790(32) 
     &~~0.309135(11)  &0.0016104(54) &~~0.309161(27)  &0.002079(13)  \\
\end{tabular}
\end{center}
\end{table}


\begin{table}[ht]
\renewcommand{\tablename}{Table}
\caption{\baselineskip=0.8cm  Bootstrap bias-corrected estimates and 
          bias-corrected standard error for the parameters
          $\alpha_u$ and $\gamma_u$ for $q=4$ and $q=5$.}
\begin{center}
\begin{tabular}{clllll}\\
\\[-0.3cm]
    &~~~~$q=4$   &             &   &~~~~$q=5$    &          \\  
$~L$&~~$\alpha_u$&~~~$\gamma_u$&~~ &~~~$\alpha_u$&~~~~$\gamma_u$ \\
\\[-0.35cm]
\hline
\\[-0.3cm]
~~8 &        &           &     & 0.402(55) &-0.2267(20)\\
~12 &        &           &     & 0.322(51) &-0.1819(16)\\
~16 &0.65(12)&-0.1897(43)&     & 0.42(11)  &-0.1552(24)\\
~20 &        &           &     & 0.479(84) &-0.1503(19)\\
~24 &        &           &     & 0.510(74) &-0.1414(23)\\
~32 &0.38(13)&-0.1650(84)&     & 0.541(93) &-0.1233(25) \\
~64 &0.44(16)&-0.058(18) &     & 0.433(65) &-0.1074(15) \\
~96 &        &           &     & 0.458(73) &-0.0970(46)\\
128 &0.36(18)&-0.125(15) &     & 0.329(40) &-0.0990(58) \\
\end{tabular}
\end{center}
\end{table}


\begin{table}[ht]
\renewcommand{\tablename}{Table}
\caption{\baselineskip=0.8cm  Bootstrap bias-corrected estimates and 
          bias-corrected standard error for the JK parameters  
          from first three zeros for $q=4$.}
\begin{center}
\begin{tabular}{ccccc}\\
\\[-0.3cm]
$~L$  &  $  a_1$ &$a_2$     & $(L,L')$  &   $a_3$ \\
\\[-0.35cm]
\hline
\\[-0.3cm]
~16   & 1.26(03) &1.574(07) & (16, 32)  &-0.00008(93) \\
~32   & 1.48(10) &1.579(16) & (32, 64)  &-0.00003(24)\\
~64   & 1.83(19) &1.591(19) &~(64, 128) &~-0.000006(59)\\
128   & 1.86(22) &1.570(18) & &\\
\end{tabular}
\end{center}
\end{table}


\begin{table}[ht]
\renewcommand{\tablename}{Table}
\caption{\baselineskip=0.8cm  Bootstrap bias-corrected estimates and 
          bias-corrected standard error for the JK parameters  
          from first four zeros for $q=5$.}
\begin{center}
\begin{tabular}{lclllcll}\\
\\[-0.3cm]
$~L$ & $a_1$   &~~~$a_2$   &~~~$A_1$  &~~~$a_3$  &~~&~$(L,L')$ &~~~$a_3$  \\
\\[-0.35cm]
\hline
\\[-0.3cm]
~~8 &1.305(28) &1.4981(78) &0.4119(18)&-0.002(25)&~~& ~(~8, 12)&-0.0003(60)\\
~12 &1.200(27) &1.4546(63) &0.3236(12)&-0.001(11)&~~&~(12, 16)&-0.0003(37)\\
~16 &1.170(40) &1.4370(81) &0.2759(14)&-0.0001(55)&~~&~(16, 20)&-0.0001(24)\\
~20 &1.161(36) &1.4290(67) &0.2451(10)&-0.0001(35)&~~&~(20, 24)&-0.0001(16)\\
~24 &1.087(30) &1.4096(60) &0.22134(77)&~0.0001(23)&~~&~(24, 32)&~0.00006(91)\\
~32 &1.045(30) &1.3949(57) &0.19069(62)&~0.0001(13)&~~&~(32, 64)&~0.00001(17)\\
~64 &0.863(41)&1.3475(82)&0.13353(76)&~0.00002(31)&~~&~(64, 96)&~0.000009(85)\\
~96&0.734(31)&1.3175(62)&0.11049(42)&~0.00001(14)&~~&~(96, 128)&~0.000007(50)\\
128 &0.666(29) &1.3002(66) &0.09687(35) &~0.000006(74)&~~&      &          \\
\end{tabular}
\end{center}
\end{table}

\begin{table}[ht]
\renewcommand{\tablename}{Table}
\caption{\baselineskip=0.8cm Partition function zeros for polyalanine.}
\begin{center}
\begin{tabular}{lcccccccc}\\
\\[-0.3cm]
$~N$~~&Re$(u_1)$ &Im$(u_1)$    & Re$(u_2)$ & Im$(u_2)$ &Re$(u_3)$ &Im
$(u_3)$ &Re$(u_4)$ &Im$(u_4)$  \\
\\[-0.35cm]
\hline
\\[-0.3cm]
~10 &0.30530(12) &0.07720(14)  &0.2823(13) &0.13820(61)&0.2459(72)&0.1851(63)
    &0.172(11)   &0.2200(71)    \\
~15 &0.356863(61)&~0.053346(39)&0.34167(60)&0.10440(59)&0.3331(48)&0.1454(28)
    &~0.3067(81) &0.1689(32)    \\
~20 &0.374016(41)&~0.042331(45)&0.36161(27)&0.08109(24)&0.3569(27)&0.1154(13)
    &~0.3336(56) &0.1470(27)    \\
~30 &0.378189(19)&~0.027167(32)&~0.37399(14)&~0.05420(27)&0.3693(11)
    &0.0804(13)  &~0.35854(63) &0.1022(43)    \\
\end{tabular}
\end{center}
\end{table}


\begin{table}[ht]
\renewcommand{\tablename}{Table}
\caption{\baselineskip=0.8cm  Bootstrap bias-corrected estimates and
          bias-corrected standard error for the parameters
          $\alpha_u$ and $\gamma_u$ for polyalanine model.}
\begin{center}
\begin{tabular}{cll}\\
\\[-0.3cm]
$~N$  &~~~$\alpha_u$ & ~~~$\gamma_u$  \\
\\[-0.35cm]
\hline
\\[-0.3cm]
~10   &-0.36(17) & -0.365(17) \\
~15   & 0.41(19) & -0.291(11) \\
~20   & 0.06(14) & -0.3229(78)\\
~30   & 0.19(14) & -0.1568(58) \\
\end{tabular}
\end{center}
\end{table}

\begin{table}[ht]
\renewcommand{\tablename}{Table}
\caption{\baselineskip=0.8cm  Bootstrap bias-corrected estimates and
         bias-corrected standard error for the JK parameters 
         for polyalanine.}
\begin{center}
\begin{tabular}{lccc}\\
\\[-0.3cm]
$~N$ & $a_1$   &  $a_2$   &  $a_3$ \\
\\[-0.35cm]
\hline
\\[-0.3cm]
~10 & 6.17(60) &1.862(46) &  0.01(14)  \\
~15 & 4.37(19) &1.664(16) &  0.014(69) \\
~20 & 3.62(26) &1.558(24) &~-0.014(98) \\
~30 & 3.54(31) &1.473(30) &~-0.007(61) \\
\end{tabular}
\end{center}
\end{table}


\newpage
\cleardoublepage

\begin{figure}[b]
\begin{center}
\begin{minipage}[t]{0.95\textwidth}
\centering
\includegraphics[angle=-90,width=0.72\textwidth]{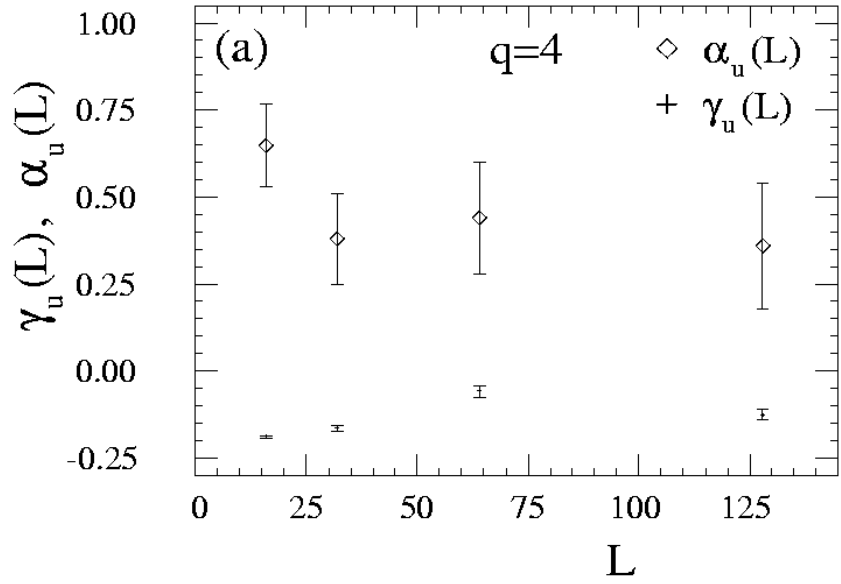}
\renewcommand{\figurename}{(Fig.1a)}
\caption{$\alpha_u(L)$ and $\gamma_u(L)$ estimates for $q=4$ Potts model
        data (a) and for $q=5$ Potts model data (b).} 
\label{Fig. 1a}
\end{minipage}
\end{center}
\end{figure}

\begin{figure}[b]
\begin{center}
\begin{minipage}[t]{0.95\textwidth}
\centering
\includegraphics[angle=-90,width=0.72\textwidth]{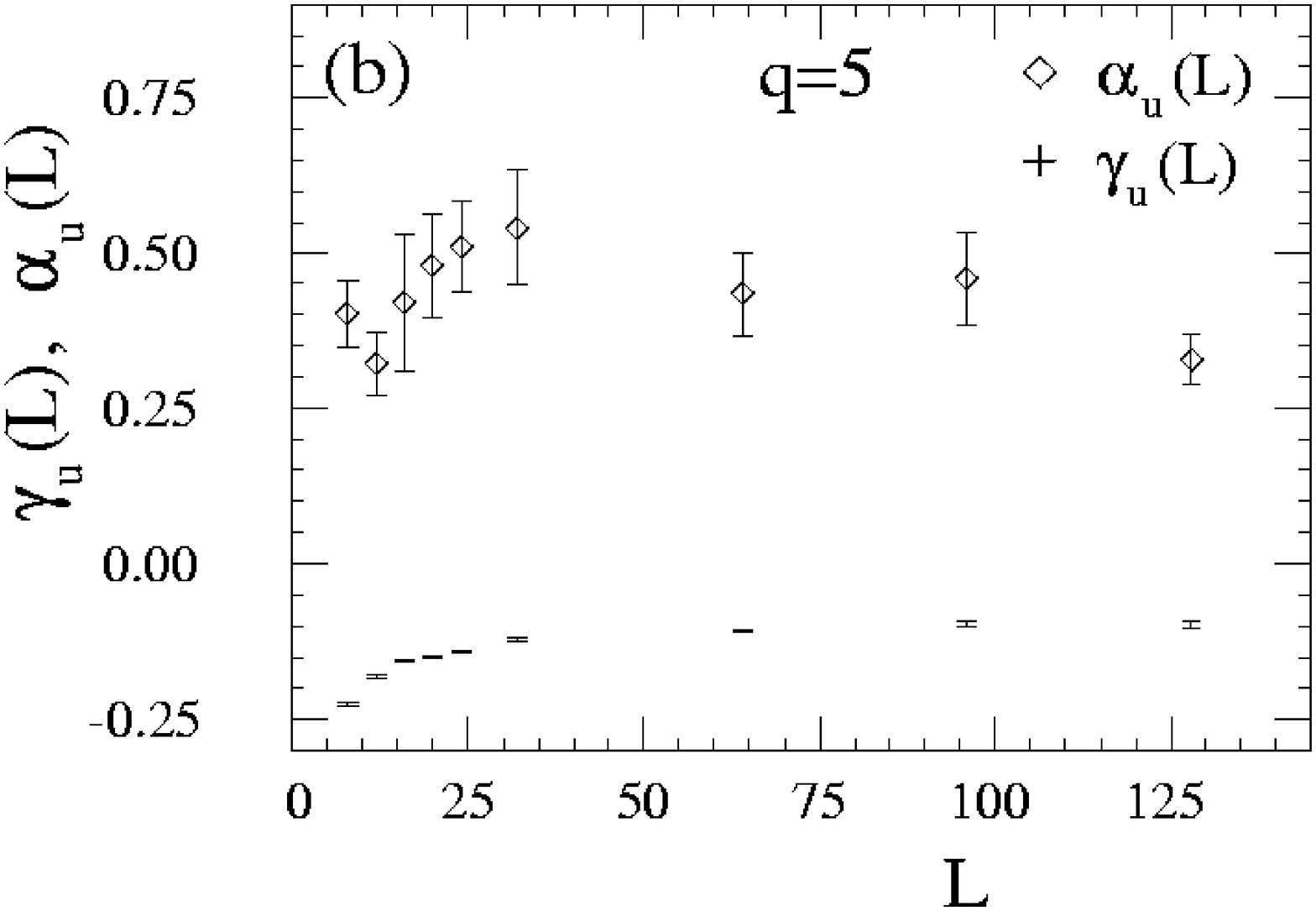}
\renewcommand{\figurename}{(Fig.1b)}
\caption{  }
\label{Fig. 1b}
\end{minipage}
\end{center}
\end{figure}

\begin{figure}[t]
\begin{center}
\begin{minipage}[t]{0.95\textwidth}
\centering
\includegraphics[angle=-90,width=0.72\textwidth]{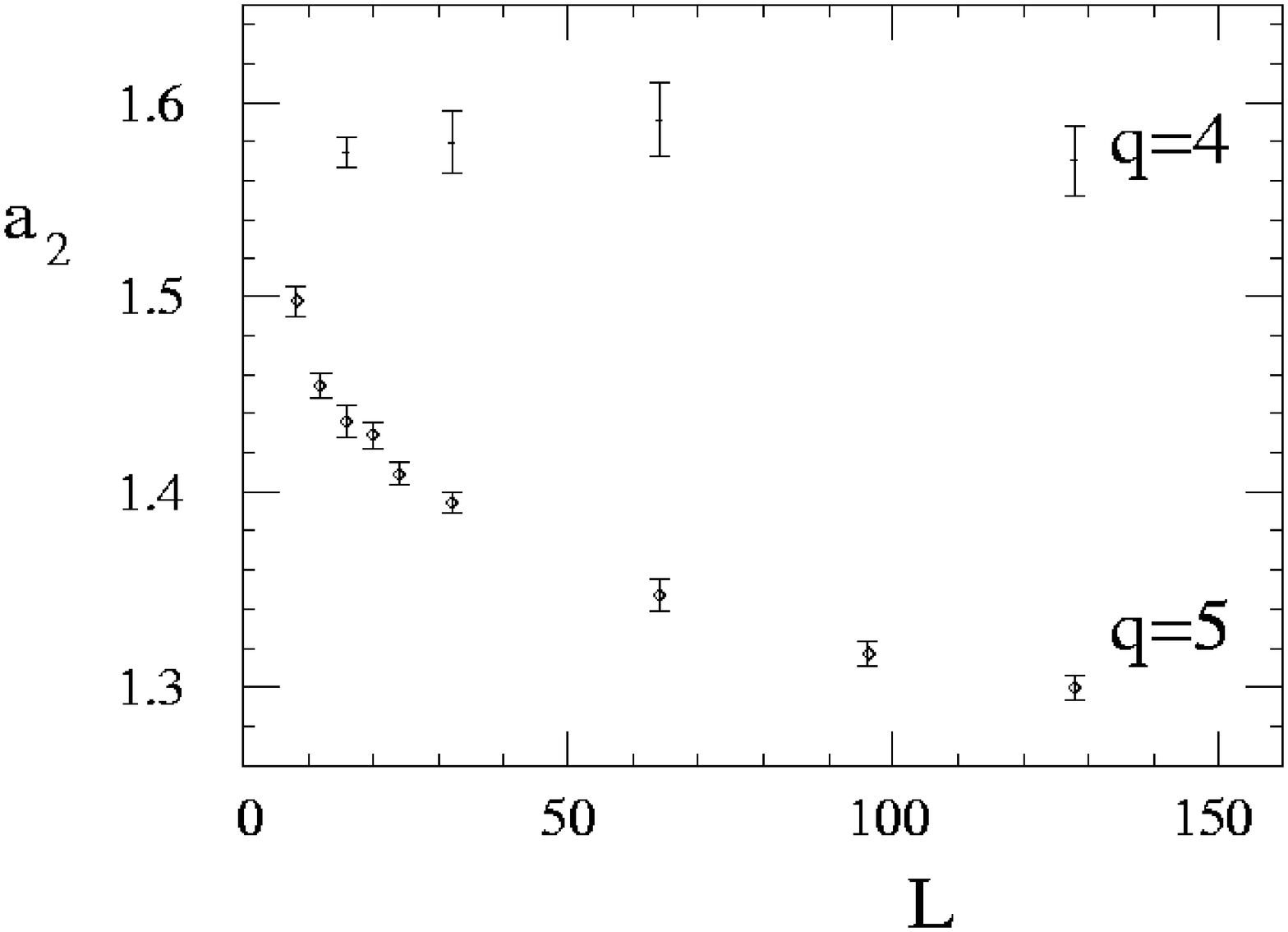}
\renewcommand{\figurename}{(Fig.2)}
\caption{The parameter $a_2(L)$ as function of system size $L$ for 
          the $q=4$ Potts model and the $q=5$ Potts model.}
\label{fig. 2}
\end{minipage}
\end{center}
\end{figure}


\begin{figure}[t]
\begin{center}
\begin{minipage}[t]{0.95\textwidth}
\centering
\includegraphics[angle=-90,width=0.72\textwidth]{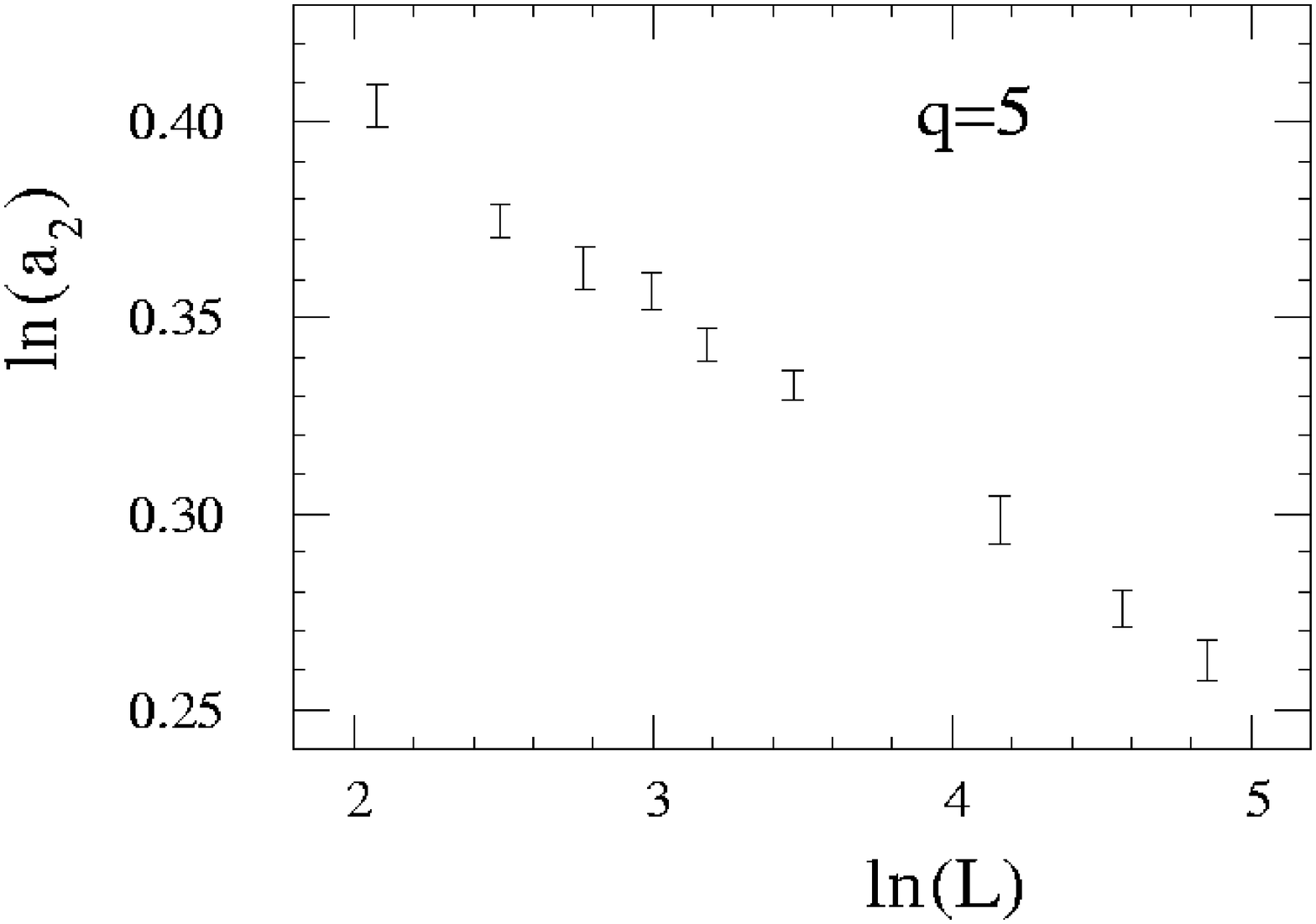}
\renewcommand{\figurename}{(Fig.3) }
\caption{The parameter $a_2(L)$ as a function of system size $L$ for
          the $q=5$ Potts model in a log-log plot.}
\label{fig. 3}
\end{minipage}
\end{center}
\end{figure}

\begin{figure}[t]
\begin{center}
\begin{minipage}[t]{0.95\textwidth}
\centering
\includegraphics[angle=-90,width=0.72\textwidth]{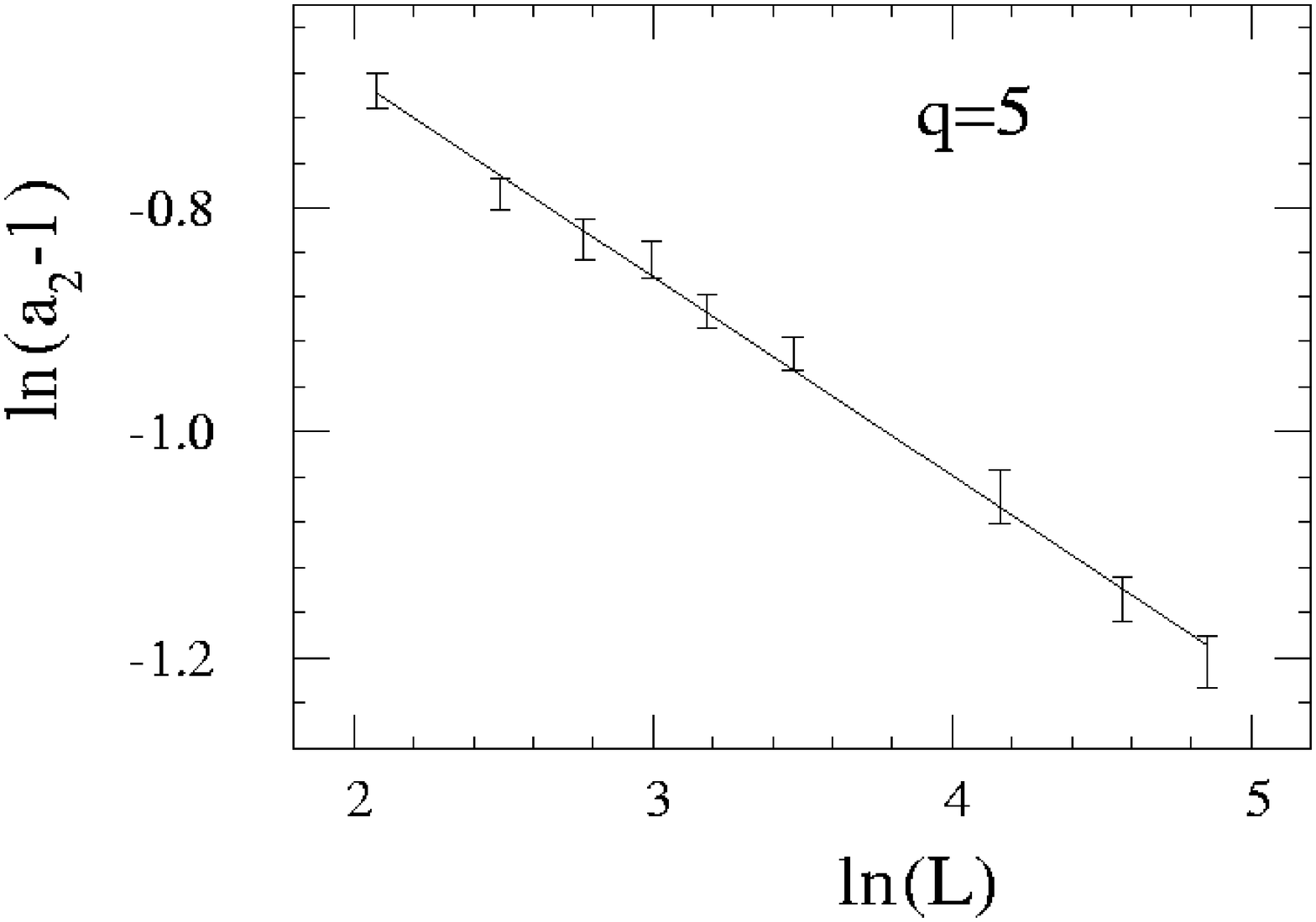}
\renewcommand{\figurename}{(Fig.4) }
\caption{$a_2(L) - 1 (=a_2(\infty))$ as a function of system size $L$
          in a log-log plot.
          The straight line through the data points is from our fit.}
\label{fig. 4}
\end{minipage}
\end{center}
\end{figure}

\begin{figure}[t]
\begin{center}
\begin{minipage}[t]{0.95\textwidth}
\centering
\includegraphics[angle=-90,width=0.72\textwidth]{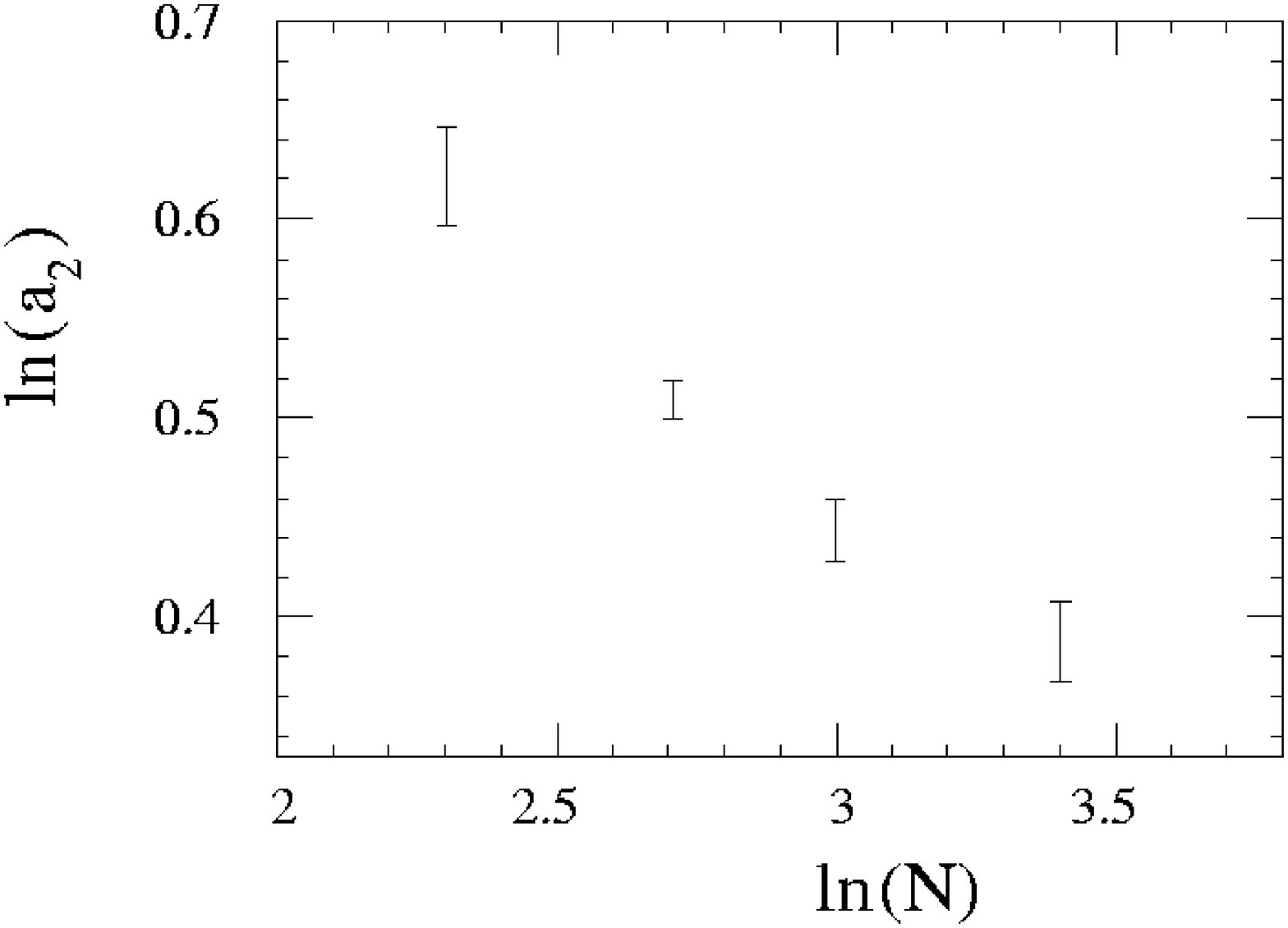}
\renewcommand{\figurename}{(Fig.5) }
\caption{The parameter $a_2(N)$ for polyalanine molecules of length $N$
          in a log-log plot.}
\label{fig. 5}
\end{minipage}
\end{center}
\end{figure}

\begin{figure}[b]
\begin{center}
\begin{minipage}[t]{0.95\textwidth}
\centering
\includegraphics[angle=-90,width=0.72\textwidth]{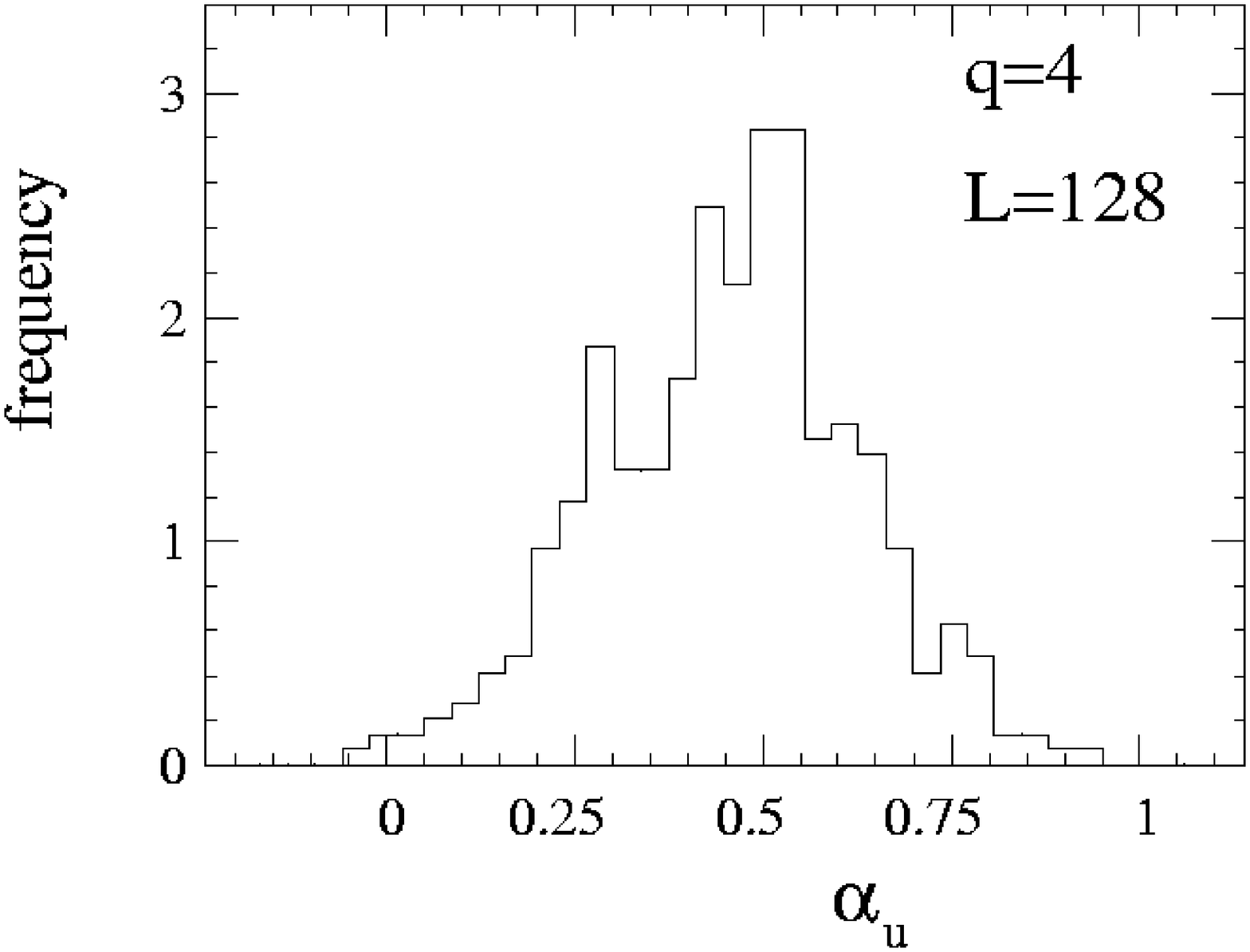}
\renewcommand{\figurename}{(Fig.6) }
\caption{Histogram of 400 bootstrap replications for
         the parameter $\alpha_u$ from four-state Potts model data.}
\label{fig. 6}
\end{minipage}
\end{center}
\end{figure}


\begin{thebibliography}{99}

\bibitem{CSSG} A. Chbihi, O. Schapiro, S. Salou and D.H.E. Gross, 
               Eur. Phys. J. A5 (1999) 251
\bibitem{Borderie}B. Borderie {\it et al.}, Phys. Rev. Lett. 86 (2001)3252.
\bibitem{Lopez}O. Lopez, Nucl. Phys. A685 (2001)246c.
\bibitem{BMH} P. Borrmann, O. M\"ulken and J. Harting, 
              Phys. Rev. Lett. 84 (2000)3511;
              O. M\"ulken, P. Borrmann, J. Harting and H. Stamerjohanns,
              Phys. Rev. A64 (2001) 013611;
              O. M\"ulken and P. Borrmann, Phys. Rev. C63 (2001) 024306.
\bibitem{Berry} A. Proykova and R.S. Berry, Z. Phys. D. {\bf 40} (1997) 215.
\bibitem{Anf} C.B.~Anfinsen, {\it Science} {\bf 181} (1973) 223.
\bibitem{Gross} D.H.E. Gross and E. Votyakov, Eur. Phys. J B15 (2000)115 ;
                D.H.E. Gross, Nucl. Phys. A681 (2001) 366c;
                D.H.E. Gross {\it http://arXiv.org/abs/cond-mat/0006203};
                D.H.E. Gross {\it http://arXiv.org/abs/cond-mat/0105313}.
\bibitem{JK}   W. Janke and R. Kenna, J. Stat. Phys. 102 (2001) 1211;
               {\it http://arXiv.org/abs/cond-mat/0103333}.
\bibitem{GR1} S. Grossmann and W. Rosenhauer, Z. Physik 207 (1967)138. 
\bibitem{GR2} S. Grossmann and W. Rosenhauer, Z. Physik 218 (1969)437; 
              S. Grossmann and W. Rosenhauer, Z. Physik 218 (1969)449.
\bibitem{Thesis} R. Villanova, Ph.D. Thesis (1991). 
                 Florida State University (unpublished).  
\bibitem{AlvesNB20} R. Villanova, N.A. Alves and B.A. Berg,
                    Nucl. Phys. B (Proc. Suppl.) 20 (1991) 665.
\bibitem{AlvesB43} N.A. Alves, B.A. Berg and R. Villanova,
                   Phys. Rev. B43 (1991) 5846.
\bibitem{Wu}  F.Y. Wu, Rev. Mod. Phys. 54 (1982)235, 
              erratum: Rev. Mod. Phys. 55 (1983)315.
\bibitem{AlvesIJMPC} N.A. Alves, J.R.D. de Felicio and U.H.E. Hansmann,
                     Int. J. Mod. Phys. C8 (1997) 1063.
\bibitem{AlvesNPB92} N.A. Alves, B.A. Berg and S. Sanielevici,
                     Nucl. Phys. B376 (1992) 218.
\bibitem{Efron} B. Efron, SIAM Rev. 21 (1979) 460.
              B. Efron and R.J. Tibshirani,{\it An introduction to 
              the Bootstrap}, Monographs on Statistics and Applied
              Probability 57 (Chapman \& Hall, 1993).
\bibitem{LandauB39} P. Peczak and D.P. Landau, 
                    Phys. Rev. B39 (1989) 11932.
\bibitem{Nau} M. Nauenberg and D.J. Scalapino, Phys. Rev. Lett. 44 (1980) 837.
              J.L. Cardy, M. Nauenberg and D.J. Scalapino, 
              Phys. Rev. B22 (1980) 2560.
\bibitem{Creswick} R.J. Creswick and S-Y Kim,
                   J. Phys.A: Math. Gen. 30 (1997) 8785.
\bibitem{Sokal} J. Salas and A.D. Sokal, J. Stat. Phys. 88 (1997) 567.
\bibitem{Caselle} M. Caselle, R. Tateo and S. Vinti, 
                  Nucl. Phys. B562 [FS] (1999) 549.
\bibitem{recipes}W. Press {\it et al.}, {\it Numerical Recipes} 
                 (Cambridge University Press, London, 1986).
\bibitem{NiOk} T. Nishino and K. Okunishi, 
                  J. Phys. Soc. Japan 67 (1998) 1492.
\bibitem{BSG} R.M.~Ballew, J.~Sabelko, and M.~Gruebele,
              Proc.~Nat.~Acad.~Sci. USA {\bf 93}, 5759 (1996).
\bibitem{Poland} D.~Poland and H.A.~Scheraga, {\it Theory of Helix-Coil
         Transitions in Biopolymers} (Academic Press, New York, 1970).
\bibitem{HO98c} U.H.E.~Hansmann and Y.~Okamoto,  
                {\it J. Chem.~Phys.} {\bf 110}, 1267 (1999), 
                                     {\bf 111} (1999) 1339(E).
\bibitem{AH99b} N.A. Alves  and U.H.E. Hansmann, 
                {Phys. Rev. Lett.}  {\bf 84} (2000) 1836.
\bibitem{KHC99c} J.P.~Kemp, U.H.E.~Hansmann and Zh.Y.~Chen,
                 {\it Eur.~Phy.~J.~B} {\bf 15}, 371 (2000).
\bibitem{EC}  M.J. Sippl, G. N{\'e}methy and H.A. Scheraga,
              {\it J.  Phys. Chem.}, {\bf 88}, 6231 (1994);
              and references therein.
\bibitem{Konf} H.~Kawai, Y.~Okamoto, M.~Fukugita, T.~Nakazawa, and
               T.~Kikuchi,   Chem. Lett. {\bf 1991}, 213 (1991);
               Y.~Okamoto, M.~Fukugita, T.~Nakazawa, and H.~Kawai,
               Protein Engineering {\bf 4}, 639 (1991).
\bibitem{MU} B.A.~Berg and T.~Neuhaus, Phys. Lett. {\bf 267}, 249 (1991).
\end{thebibliography}
\end{document}